# Structural, Optical and Magnetic Properties of Superparamagnetic $Fe_3O_4@TiO_2$ and $Fe_3O_4@SiO_2@TiO_2$ Core–Shell Nanostructures

Wagner Henrique, Gamas.[a*] Rosana, R.Rangel[a]. Grécia Alejandra, Gomez-Iriarte[b] Sergio, Seabra.[a] Pablo, L.Bernardo.[a*]

[*]Corresponding authors
E-mail addresses: henriquevnno@pq.uenf.br ( WH Gamas), pablolb@uenf.br (PL Bernardo).

[a] Universidade Estadual do Norte Fluminense Darcy Ribeiro (UENF)
[b] LABNANO, Centro Brasileiro de Pesquisas Físicas (CBPF)

## Highlights

- Multifunctional magnetic core and photo-responsive shell nanostructure were synthesized by the sol-gel method;
- XRD and FTIR data confirmed a thick tetragonal titanium dioxide phase shell without spurious phases;
- TEM and HR-SEM results exhibit well-dispersed particles in the range of nanometric scale;
- Magnetic data confirmed superparamagnetic behavior in all samples.
- Diffuse reflectance spectroscopy confirmed redshift from nanostructures absorption in comparison to pure titanium dioxide.

## Abstract

Tailoring core-shell nanostructures using iron oxide core ($Fe_3O_4$) and titanium dioxide shell ($TiO_2$) offer a promising multifunctional platform for multimodal cancer therapies. $TiO_2$ shell exhibits photo-responsive optical properties suitable for optical imaging and phototherapies, while $Fe_3O_4$ magnetic core enables magnetic resonance imaging (MRI) and magnetic hyperthermia therapy (MHT). This present work reports synthesis and characterization from core-shell $Fe_3O_4@TiO_2$ and $Fe_3O_4@SiO_2@TiO_2$ nanostructures. Structural and physical properties were evaluated using X-ray Diffraction (XRD), Fourier-Transform Infrared Spectroscopy (FTIR), Transmission Electron Microscopy (TEM), High-Resolution Scanning Electron Microscopy (HRSEM), Diffuse UV-Vis Reflectance Spectroscopy (DRS) and SQUID magnetometry. This work highlights the potential use of magnetic (core) and optically activated (shell) nanostructures for multimodal therapies assisted by magnetic hyperthermia.

## 1. Introduction

In magnetic hyperthermia therapy (MHT), superparamagnetic nanoparticles generate localized heat under external AC magnetic fields through Néel and Brownian relaxations in single-domain regimes, exhibiting zero remnant magnetization and coercivity [1,2]. Advancing MHT depends on optimizing field devices, targeted delivery carriers, clinical protocols and engineered magnetic nanomaterials. Magnetite ($Fe_3O_4$) stands out as promising magnetic agent for MHT due to its inverse spinel structure and dual $Fe^{2+}$/$Fe^{3+}$ valence states [2]. Integrating magnetic cores with different stimuli shells (ultrasound, magnetic, light or ph) creates multifunctional nanostructures that may target tumors in more than one way [3]. For example, under UV radiation, FDA-approved anatase titanium dioxide ($TiO_2$) shell excites valence electrons to the conduction band, generating surface electron-hole pairs (e-/h+) that may transfer charges, ultimately triggering selective cancer cell apoptosis and antimicrobial effects [3-5]. This work investigates $Fe_3O_4$@$TiO_2$ and $Fe_3O_4$@$SiO_2$@$TiO_2$ nanostructures using XRD, FTIR, TEM, HR-SEM, DRS, and SQUID magnetometry to demonstrate their potential in hyperthermia-assisted multimodal cancer therapies.

## 2. Experimental

### 2.1 Core-shell *$Fe_3O_4$@$TiO_2$* and *$Fe_3O_4$@$SiO_2$@$TiO_2$* synthesis

Core-shell $Fe_3O_4$@$SiO_2$ nanoparticles were prepared by modified Stöber method [6]. Typically, an $Fe_3O_4$ amount was dispersed in absolute ethanol via probe sonication for 50 min. Ultrapure water and $NH_4OH$ were added under magnetic stirring, followed by tetraethyl orthosilicate (TEOS). The mixture was maintained under constant stirring for 24 h at room temperature. The $Fe_3O_4$@$SiO_2$ product was isolated via magnetic separation, washed with absolute ethanol, and dried at room temperature. $TiO_2$ were coated onto the cores via modified sol-gel route [7]. The precursors titanium isopropoxide (TTIP), 2-propanol, glacial acetic acid (GAA), ultrapure water, and ethylene glycol (EG) were purchased from Sigma-Aldrich. Briefly, Solution A was prepared by dissolving TTIP in a 2-propanol and GAA mixture under magnetic stirring. Solution B, containing 2-propanol, ultrapure water, and EG, was then added dropwise to Solution A at 90 °C under continuous stirring. Resulting gel was transferred to a muffle furnace and dried at heating rate of 10 °C/min. Finally, the dried product was ground and calcined at 500 °C for 4 h. Uncoated $TiO_2$ sample were prepared for comparative XRD analysis.

## 3. Results and discussion

### 3.1. Structural and morphological characterization

XRD patterns were acquired using a Panalytical Empyrean diffractometer with Cu tube with λ= 1,5406 Å . The **Figure 3.1.1(a–d)** presents Rietveld refinement profiles for the undoped $TiO_2$, pure $TiO_2$, $Fe_3O_4$@$TiO_2$, and *$Fe_3O_4$*@$SiO_2$@$TiO_2$ samples, respectively. Undoped $TiO_2$ sample exhibited cubic $Fe_3O_4$ structure (space group Fd-3m, COD No. 00-900-5842) with lattice parameters a=b=c=8.3744Å, α=β=γ=$90^0$, and V=587.30Å$^3$ (GoF = 1.2; $\chi^2$ = 1.46). Pure $TiO_2$, $Fe_3O_4$@$TiO_2$, and $Fe_3O_4$@$SiO_2$@$TiO_2$ samples displayed tetragonal crystalline arrangement (space group I4/amd, COD No. 00-900-9086) with parameters a=b=3.78Å, c=9.50Å, α=β=γ=$90^0$, and V=135.98Å$^3$. Refinements indices (GoF ≤ 1.5 and $\chi^2$ ≤ 2.28), indicating excellent agreement between calculated and experimental profiles. Notably,

core–shell samples showed no reflections corresponding to iron oxide or secondary phases, confirming the successful formation of $TiO_2$ shell.

**Figure 3.1.1: Rietveld refinement data from (a) un-doped titanium dioxide, (b) pure titanium dioxide, (c) magnetite@titanium (d) magnetite@silica@titanium.**

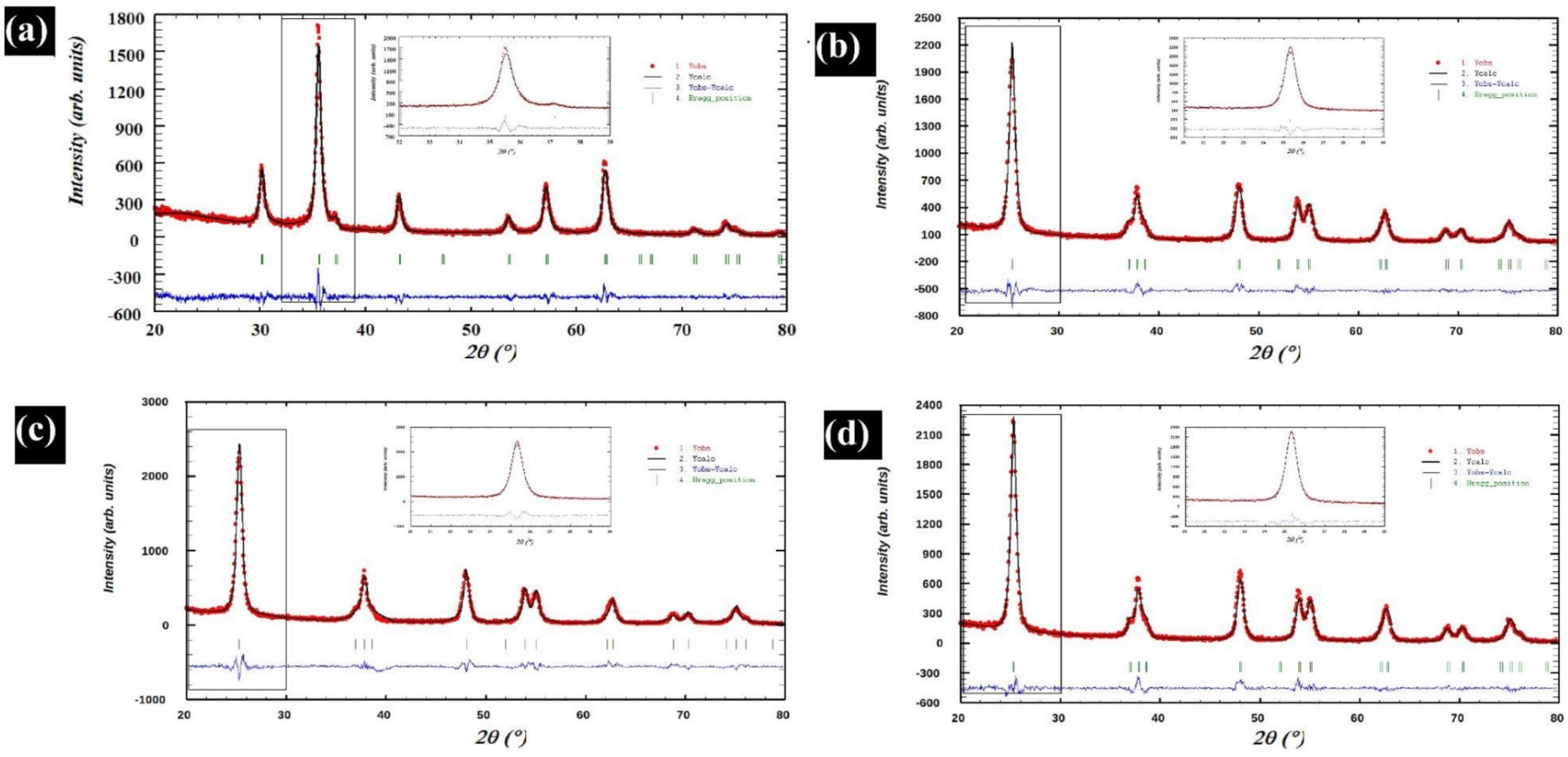

Average crystallite sizes and lattice strains were evaluated via Scherrer and modified Williamson-Hall (W-H) methods using PCrystalX software [8]. For the uncoated $Fe_3O_4$ core, crystallite sizes were 14.97 ± 0.18 nm (Scherrer, Fig. 3.1.2a) and 14.98 ± 0.01 nm (W-H, Fig. 3.1.2b). Upon $TiO_2$ shell formation, average sizes decreased to 11.77 ± 0.09 nm (Scherrer, Fig. 3.1.2c) and 12.59 ± 0.01 nm (W-H, Fig. 3.1.2d). This reduction suggests that the core restricts $TiO_2$ growth, inducing tensile strain at the interface and unit cell contraction [9]. FTIR spectra (Fig. 3.1.2e) confirmed complete precursor combustion with no impurities. The bare core showed the characteristic Fe–O stretching near 570 $cm^{-1}$ [6,10]. In $Fe_3O_4$@$TiO_2$ and $Fe_3O_4$@$SiO_2$@$TiO_2$, this band was masked by Ti–O/Ti–O–Ti absorption (400–800 $cm^{-1}$) [4-6,10], confirming core encapsulation. Silica coating introduced an Si–OH asymmetric stretching band near 996 $cm^{-1}$ [4-6,11]. Finally, O–H and H–O–H bands (~3400 $cm^{-1}$ and 1600 $cm^{-1}$) decreased in intensity after coating, indicating physical shielding of core surface groups by $SiO_2$ and $TiO_2$ shells.

**Figure 3.1.2: Estimated crystallite sizes for undoped titanium dioxide (a) by Scherrer, (b) by modified Williamson-Hall, and for titanium dioxide samples (c) by Scherrer, (d) by modified Williamson-Hall, and (e) FT-IR spectrum for all samples.**

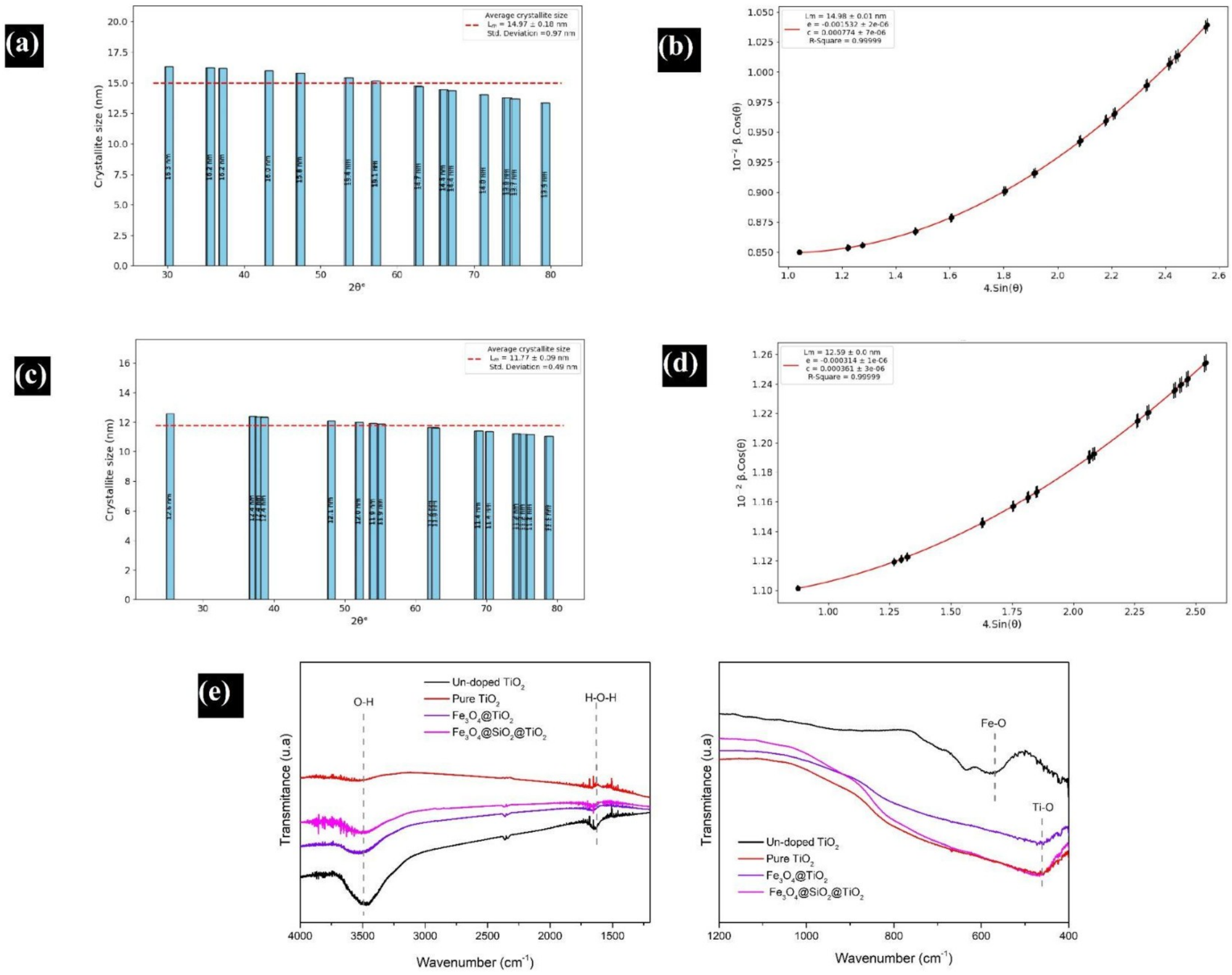

Morphological studies were examined via TEM (Jeol JEM-1400 Plus), HR-SEM (Jeol 7100FT), and particle sizes estimated using MagMicros software [12]. In the $Fe_3O_4$@$TiO_2$ sample, TEM and HR-SEM images (**Fig. 3.1.3a–b**) reveal highly agglomerated clusters (~195.65 nm) featuring a nanoflake morphology but poor visual distinction between the core and shell phases. Conversely, the $Fe_3O_4$@$SiO_2$@$TiO_2$ nanocomposites (**Fig. 3.1.3c–d**) exhibit a significantly reduced average size (~67.73 nm), combining a well-defined spherical core-shell boundary with excellent dispersibility and low cluster agglomeration. This pronounced morphological disparity is driven by the intermediate $SiO_2$ layer. As supported by FTIR data, silica acts as both crosslinking and electrostatic templating agent [5, 6, 11]. Without this intermediate layer, the $Fe_3O_4$@$TiO_2$ structure undergoes unconstrained nucleation and crystal growth, yielding dense, large-scale particulate clusters. The incorporation of the $SiO_2$ matrix establishes defined growth boundaries for the secondary phase and increases interparticle electrostatic repulsion, which effectively restricts the final particle size and ensures high dispersion stability.

**Figure 3.1.3: Core-shell morphologies in (a) TEM, (b) HR-SEM from magnetite@titanium and from magnetite@silica@titanium in (d) TEM and (d) HR-SEM.**

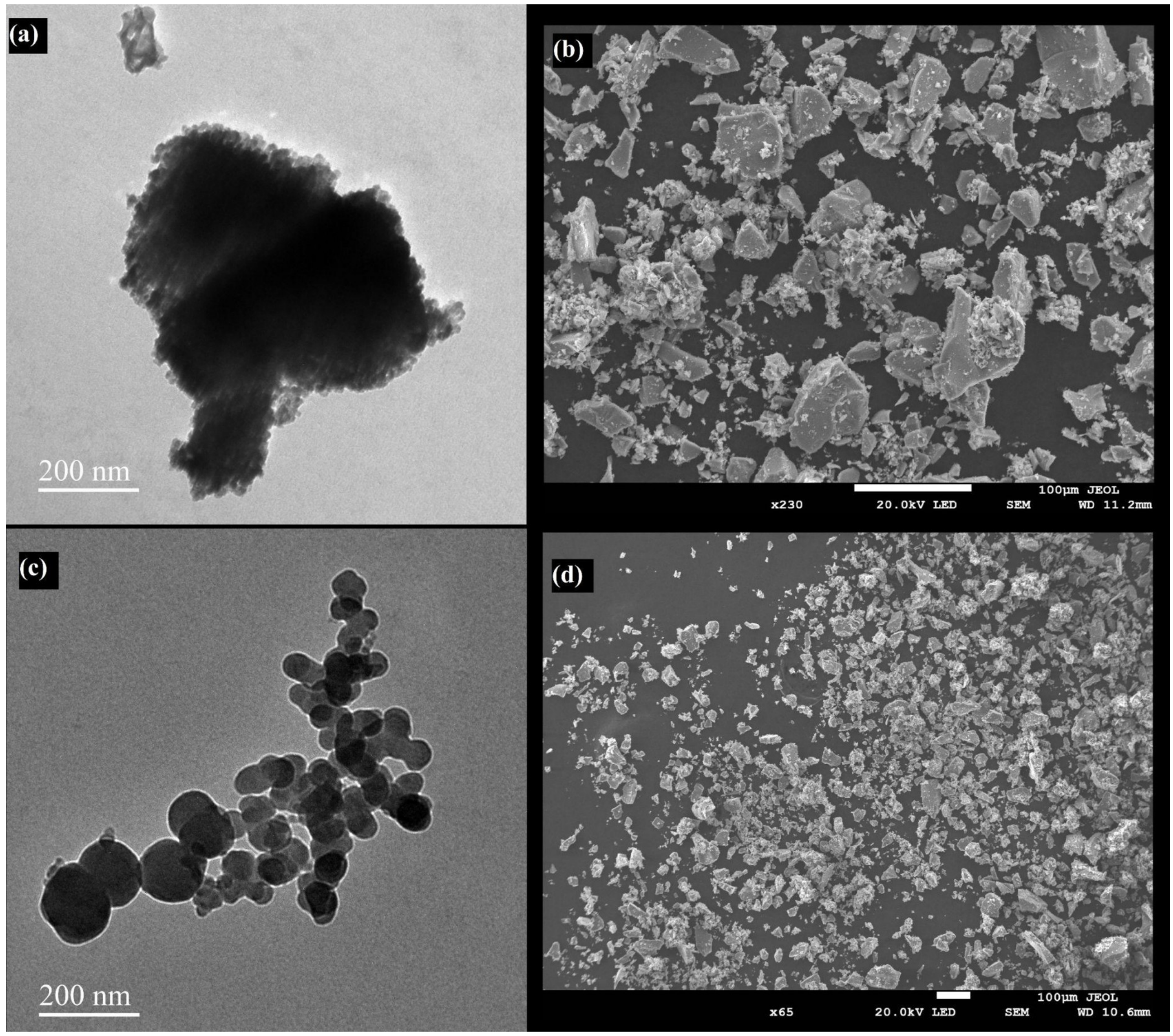

**3.2. Optical and magnetic measurements**

UV-Vis diffuse reflectance spectra (UV-2600i, Shimadzu) and corresponding indirect Tauc plot ($E_g$) are presented in **Figure 3.3.2(a-d).** UV-Vis DRS and Tauc plots revealed distinct indirect band gaps ($E_g$) equal to 2.2 eV for bare $Fe_3O_4$ and 3.32 eV for pure $TiO_2$. The $Fe_3O_4$@$TiO_2$ composite showed a blueshift in $E_g$ = 3.41 eV, driven by interfacial Fe–Ti defect states. Conversely, $Fe_3O_4$@$SiO_2$@$TiO_2$ recovered a sharp absorption edge ($E_g$=3.37 eV) similar to pure $TiO_2$. This occurs because the intermediate $SiO_2$ layer acts as an optical insulator, preventing direct charge transfer to the $Fe_3O_4$ core and preserving $TiO_2$ optical properties [5,6]. SQUID magnetometry presented in **Figure 3.3.2(e-f)** revealed bare $Fe_3O_4$ with superparamagnetic behavior and saturation magnetization (Ms) ~60 emu/g. Although both $Fe_3O_4$@$TiO_2$ and $Fe_3O_4$@$SiO_2$@$TiO_2$ nanostructures exhibited superparamagnetism, their Ms significantly decreased. In $Fe_3O_4$@$TiO_2$, this reduction stems from $Fe^{3+}$- $Ti^{4+}$ interactions disrupting core spin alignment and the diamagnetic $SiO_2$ layer further attenuates magnetization [5,6]. Additionally, high surface-to-volume ratios in these core-shell systems increase interfacial lattice disorder, inducing surface spin canting and forming a non-magnetically layer that suppresses Ms.

**Figure 3.2.1: Optical results from Diffuse UV VIS Reflectance and indirect Tauc plot from bare Fe3O4 in (a), pure TiO2 in (b), $Fe_3O_4@TiO_2$ in (c) and $Fe_3O_4@SiO_2@TiO_2$ in (d), magnetic hystersis loop from bare Fe3O4 in (e), $Fe_3O_4@TiO_2$ and $Fe_3O_4@SiO_2@TiO_2$ in (f).**

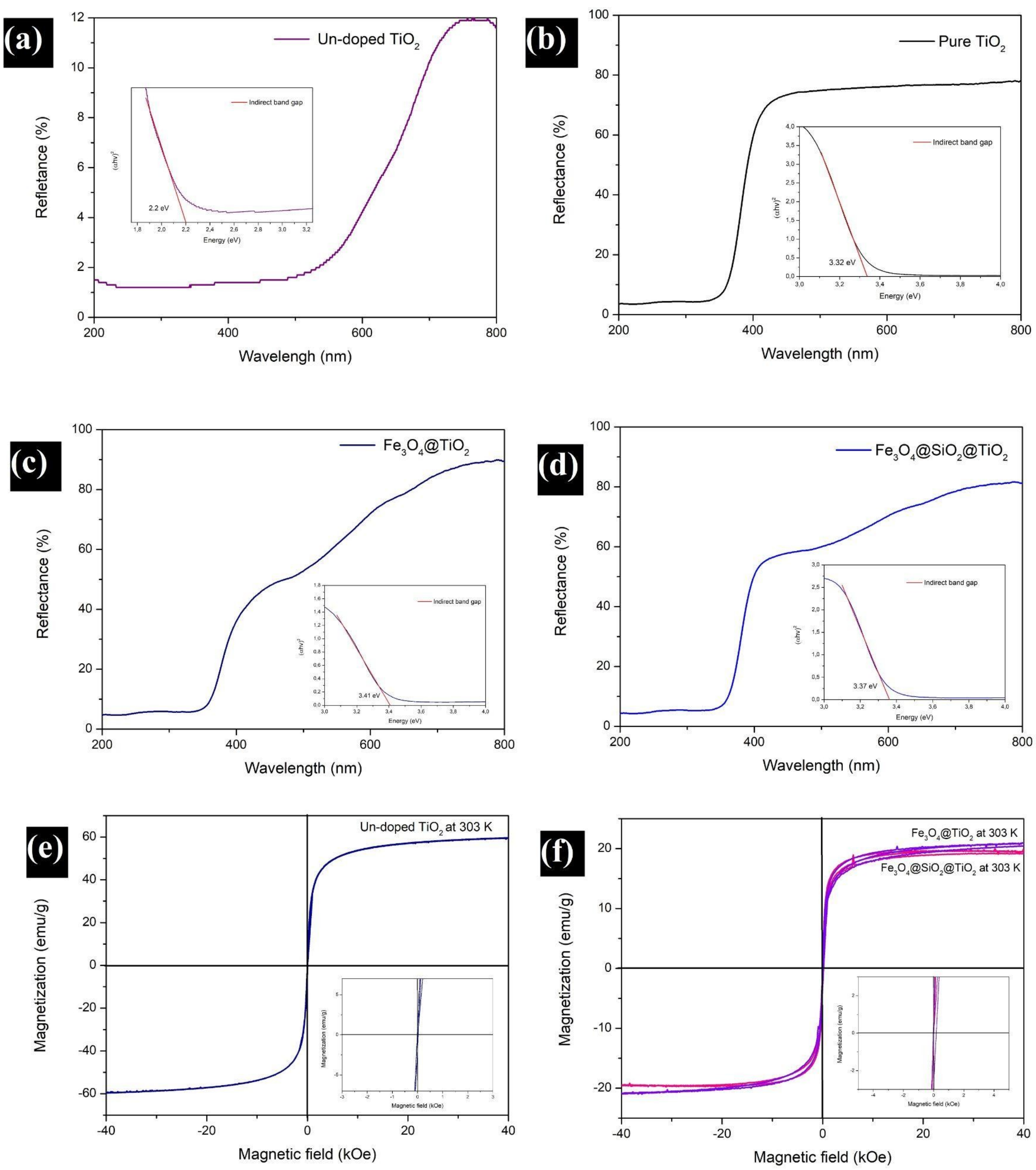

## 4. Conclusion

In summary, bare $Fe_3O_4$, pure $TiO_2$, $Fe_3O_4@TiO_2$ and $Fe_3O_4@SiO_2@TiO_2$ core-shell nanostructures were successfully synthesized and comprehensively characterized via XRD, TEM, HR-SEM, FTIR, and DRS techniques. $TiO_2$ shell formation successfully integrates optical and magnetic functionalities into a single platform, highlighting its strong potential for drug delivery and magnetic hyperthermia-assisted therapies. Future investigations involving magnetic and optical calorimetry assays, alongside comprehensive in vitro and in vivo evaluations, will further optimize this multimodal system for biomedical applications.

**CRediT authorship contribution statement**

**Gamas, HW:** Writing – review & editing, Writing – original draft, Methodology, Investigation, Formal analysis, Data curation, Conceptualization. **Rangel. RR:** Investigation, Formal analysis. **Grécia:** Investigation. **Seabra. S:** Investigation. **Bernardo. L P:** Funding acquisition, Formal analysis, Conceptualization.

**Declaration of competing interests**

The authors declare that they have no known competing financial interests or personal relationships that could have appeared to influence the work reported in this paper.

**Acknowledgements**

This study was financed by **Coordenação Aperfeiçoamento de Pessoal de Nível Superior (CAPES)** scholarship, **Fundação Carlos Chagas Filho de Amparo à Pesquisa do Estado do Rio de Janeiro (FAPERJ)**, SEI Process SEI-260003/015433/2021, SEI-260003 / 001188/ 2023, **PAPIC-UEN** SEI-260002/007082/2024 and **Conselho Nacional de Desenvolvimento Científico e Tecnológico (CNPq)** 402448/2021-9 and 07802/2022-3.

**References**

[1] CASTELO-GRANDE, Teresa et al. Economic and accessible portable homemade magnetic hyperthermia system. Materials, v. 17, n. 10, p. 2279, 2024.
[2] SPOIALĂ, Angela et al. Magnetite-silica core/shell nanostructures: from surface functionalization towards biomedical applications. Applied Sciences, v. 11, n. 22, p. 11075, 2021.
[3] HAGHIGHI, H. et al. An amplified sonodynamic therapy by a nanohybrid of titanium dioxide-gold-polyethylene glycol-curcumin: Ultrasonics Sonochemistry, v. 102, p. 106747, 2024.
[4] RAJABATHAR, Jothi Ramalingam et al. Enhanced photocatalytic activity of magnetite/titanate (Fe3O4/TiO2) nanocomposite for methylene blue dye degradation under direct sunlight. Optical Materials, v. 148, p. 114820, 2024.
[5] MAZHARI, Mohammad-Peyman et al. Development and application of multifunctional Fe3O4/SiO2/TiO2/Cu nanocomposites for sustainable water treatment. Journal of Sol-Gel Science and Technology, v. 110, n. 1, p. 156-168, 2024.
[6] KHALID, A. et al. Fe3O4 nanoparticles and Fe3O4@ SiO2 core-shell: synthesize, structural, morphological, linear, and nonlinear optical properties. Journal of Alloys and Compounds, v. 947, p. 169639, 2023.
[7] ADEDOKUN, O. et al. Structural, optical and magnetic studies of sol-gel synthesized Mg-doped pure anatase TiO2 nanoparticles for spintronic and optoelectronics applications. Physica b: Condensed Matter, v. 667, p. 415199, 2023.
[8] Bernardo, Pablo L. "PCrystalX--Web Application." *Xiv:2204.14072* (2022).
[9] CULLITY, B. Elements of. X-ray! Jiffraction, 1978.

[10] DEMIN, Alexander M. et al. Magnetic-responsive doxorubicin-containing materials based on $Fe_3O_4$ nanoparticles with a $SiO_2$/PEG shell and study of their effects on cancer cell lines. International journal of molecular sciences, v. 23, n.16, p. 9093, 2022.
[11] WANG, Lei et al. Surface-modified $TiO_2$@ $SiO_2$ nanocomposites for enhanced dispersibility and optical performance to apply in the printing process as a pigment. ACS omega, v. 8, n. 22, p. 20116-20124, 2023.
[12] Bernardo, Pablo L. "https://magnano.uenf.br/magmicros"